\documentclass[aps,prl,showpacs,twocolumn,amsmath,amssymb]{revtex4-1}
\usepackage{graphicx,bbm}

\newcommand{\be}{\begin{equation}}
\newcommand{\ee}{\end{equation}}

\newcommand{\ket}[1]{{\vert #1 \rangle}}

\newcommand{\ii}{ {\rm i} }
\newcommand{\dd}{ {\rm d} }

\newcommand{\RaR}{\mathbb{R}}

\begin{document}

\title{Eigenvalue statistics as indicator of integrability of non-equilibrium density operators}
\author{Toma\v{z} Prosen and Marko \v Znidari\v c}
\affiliation{Department of Physics, FMF,  University of Ljubljana, Jadranska 19, 1000 Ljubljana, Slovenia}

\date{\today}

\begin{abstract}
We propose to quantify the complexity of non-equilibrium steady state density operators, as well as of long-lived Liouvillian decay modes, in terms of level spacing distribution of their spectra.
Based on extensive numerical studies in a variety of models, some solvable and some unsolved, we conjecture that integrability of density operators
(e.g.,  existence of an algebraic procedure for their construction in finitely many steps) 
is signaled by a Poissonian level statistics, whereas in the generic non-integrable cases one finds level statistics of a Gaussian unitary ensemble of random matrices. 
Eigenvalue statistics can therefore be used as an efficient tool to identify integrable quantum non-equilibrium systems.
\end{abstract}

 
\maketitle

{\em Introduction.--}  
Random matrix theory (RMT) \cite{mehta} is one of the most abstract yet successful models of statistical physics which is capable of universally describing such diverse phenomena in nature and society as quantum chromodynamics \cite{verbaarschot}
 and stock exchange volatility \cite{guhr}. In general terms, RMT characterizes universal features of a certain phenomenon based on statistical correlations between the eigenvalues of a Hermitian matrix which describes the problem, be it the system Hamiltonian in a typical state basis or the covariance matrix of stocks in a portfolio. RMT then {\em explains} these eigenvalue correlations in terms of those of a probabilistic ensemble of random Hermitian matrices.
 
The so-called {\em quantum chaos conjecture} (QCC) \cite{berry,CGV81,BGS84} provided a deep connection between eigenvalue correlations of quantum Hamiltonians of non-linear single (or few-) particle problems and algorithmic complexity
of the underlying classical trajectories. Namely, it has been shown \cite{Muller} that dynamics where all classical trajectories are {\em chaotic}, i.e. exponentially unstable, results in RMT spectral fluctuation of the corresponding quantum Hamiltonian. For
Liouville integrable systems on the other side, following the argument by Berry and Tabor \cite{BT77}, the existence of a complete set of integrals of motion resulting in a full set of quantum numbers prohibits any statistical correlations in the quantum spectra and renders the corresponding level statistics {\em Poissonian}. Similarly,  based on observations \cite{poilblanc} it has been suggested that simple many-body quantum Hamiltonians that do not have classical limit possess Poissonian or RMT level statistics whenever they are integrable or strongly non-integrable, respectively.
As there is no systematic algorithmic method by which one can establish whether a certain system is integrable, i.e. exactly solvable or not, the level statistics has become a standard empirical indicator of integrability. It has been corroborated by
a vast amount of numerical and experimental data \cite{hjs}. 

QCC describes the situation of {\em closed} quantum systems. In equilibrium, the density operator is given by the Gibbsian $\rho_{\rm eq} = Z^{-1}\exp(-\beta H)$, i.e., a mixture of eigenstates of the Hamiltonain $H$, $H\ket{E_n}=E_n\ket{E_n}$, with probabilities $p_n = Z^{-1} e^{-\beta E_n}$. Since smooth, monotonous transformation $\rho \to \log\rho$ does not change the local level correlations, one can rephrase the old problem of level statistics for closed system Hamiltonians in terms of level statistics of the corresponding equilibrium density operator and formulate QCC for $\rho_{\rm eq}$.

In {\em open} quantum systems, however, the evolution of the density operator is given in terms of a master equation with the Liouvillian generator that contains both, the Hamiltonian and the dissipative terms, the latter coming from the interaction between the system and the environment. Within the Markovian approximation such evolution is given in terms of Lindblad equation \cite{lindblad}
\begin{equation}
\frac{\dd}{\dd t}\rho = \hat{\cal L}\rho := -\ii [H,\rho] + \hat{\cal D}\rho,
\end{equation}
with $\hat{\cal D}\rho :=\sum_\mu 2L_\mu \rho L_\mu^\dagger - \{L^\dagger_\mu L_\mu,\rho\}$ being the quantum dissipation fully specified by a set of quantum-jump (Lindblad) operators $L_\mu$ describing the incoherent processes in the evolution.
The positive semi-definite Hermitian operator $\rho(t)$ describes the quantum relaxation process from some initial state $\rho(0)$ to the steady state  $\rho_0=\rho(t\to\infty)$, satisfying $\hat{\cal L}\rho_0=0$.

In this Letter we formulate the QCC  for {\em non-equilibrium} density operator, say for {\em non-equilibrium steady state} (NESS) $\rho_0$, or even {\em Hermitian decay modes} (HDM), i.e. right eigenoperators of $\hat{\cal L}$ with {\em real} \cite{note} eigenvalues
$\Lambda_m$, $\hat{\cal L}\rho_m = \Lambda_m \rho_m$, where $\Lambda_0=0$. We consider level statistics of NESS and HDMs for several models of open quantum spin chains with boundary Lindblad driving and find, quite remarkably, that the former is Poissonian for all,  interacting and non-interacting cases that are exactly solvable, i.e. for which we can write $\rho_0$ explicitly in terms of a matrix product ansatz.

For models, for which already the bulk Hamiltonian $H$ is non-integrable we find, consistently, that level statistics of NESS and HDMs is described by a Gaussian Unitary Ensemble (GUE) of complex Hermitian random matrices (due to lack of time-reversal symmetry in generic non-equilibrium situations). We find GUE level statistics also for several models with integrable $H$ but for which dissipative boundary conditions break integrability \cite{marcin}. This leads us to a generalization of QCC to non-equilibrium density operators. 

In the last 30 years many solvable master equations describing classical non-equilibrium models have been discovered, in particular among lattice gas models \cite{blythe}. Exactly solvable quantum many-body master equations though are only beginning to emerge, with so far only a handful of examples, namely, quasi-free (quadratic) fermionic \cite{3Q,karevski,clark}, or bosonic \cite{3Qb}, systems with linear, or Hermitian quadratic \cite{eisler,temme,horstmann,dephasing}, noise (Lindblad) operators, and maximally boundary driven $XXZ$ chains \cite{new}. 
One of the main difficulties is in the first place identifying promising candidates of solvable quantum master equations.
Crtierion suggested in this Letter, namely the generalized QCC, could be found very useful in this respect.  For instance, in our study we find Possonian level statistics for the $XXZ$ spin $1/2$ chain at large anisotropy $\Delta$, indicating a possibility of a yet unknown exact solution for NESS in the asymtotic regime $|\Delta| \gg 1$. This could be particularly interesting as the model exhibits diffusive spin transport in this regime \cite{pz,robin,markonew}.

{\em The Models and the Method.--} We shall demonstrate our conjecture on a number of one-dimensional spin $1/2$ systems that are driven at chain boundaries and optionally exhibit a bulk dephasing. All can be described by the $XXZ$ type of Hamiltonian, $H=\sum_{j=1}^{n-1} (\sigma_j^{\rm x} \sigma_{j+1}^{\rm x} +\sigma_j^{\rm y} \sigma_{j+1}^{\rm y}+\Delta \sigma_j^{\rm z}\sigma_{j+1}^{\rm z})+\sum_{j=1}^n b_j \sigma_j^{\rm z}$, for a chain of $n$ sites. The dissipator $\hat{\cal D}=\hat{\cal D}^{\rm driv}+\gamma\,\hat{\cal D}^{\rm deph}$ is composed of a non-equilibrium driving part $\hat{\cal D}^{\rm driv}$ that acts on the first and the last spin and is described by $4$ Lindblad operators, $L_1=\sqrt{\Gamma(1-\mu+\bar{\mu})}\,\sigma^+_1,\, L_2=\sqrt{\Gamma(1+\mu-\bar{\mu})}\, \sigma^-_1$ at the left end and $L_3=\sqrt{\Gamma(1+\mu+\bar{\mu})}\,\sigma^+_n,\, L_4=\sqrt{\Gamma(1-\mu-\bar{\mu})}\, \sigma^-_n$ at the right end, and of a dephasing $\gamma\,\hat{\cal D}^{\rm deph}$ described by one Lindblad operator at each site, $L^{\rm deph}_j=\frac{1}{\sqrt{2}}\sigma^{\rm z}_j,\,\, j=1,\ldots,n$. Relevant $XXZ$ chain parameters are the anisotropy $\Delta$ and the external magnetic field $b_j$. For a homogeneous field the system is integrable \cite{kbi93}, with the anisotropy changing the magnetization transport properties from ballistic for $|\Delta|<1$ to diffusive for $\Delta>1$ (in the absence of the field, $b_j\equiv 0$). A staggered magnetic field renders the system quantum chaotic \cite{thermalization}. Dissipative parameters, describing the influence of environmental degrees of freedom, are the dephasing strength $\gamma$ (a nonzero value causes the system to become diffusive), the coupling strength $\Gamma$ (its precise value is inessential) and two driving parameters, the driving strength $\mu$ that determines how far from equilibrium we are ($\mu=0$ causes an infinite-temperature equilibrium state) and $\bar{\mu}$ that determines the average magnetization.

The above class of systems includes solvable as well as non-solvable out-of-equilibrium models. We shall first study spectral statistics of NESSs and at the end consider also HDMs. In each case we calculate the NESS $\rho_0$ (or a HDM) numerically exactly by using either an explicit solution, if it is known, or, by numerically finding the eigenvector of the Liouvillian $\hat{\cal L}$ using the  Arnoldi method. For veryfing generalized QCC we have to assess as large a system as possible. Exponentially growing dimension of $\rho_0$ limits us to systems of about $n=20$ sites for solvable models; for non-solvable systems the limiting factor is actually not the diagonalization of $\rho_0$ (being of dimension $2^n$) but rather solving for NESS, $\hat{\cal L}\rho_0=0$ (a set of $4^n$ linear equations). In all systems studied the total magnetization $Z=\sum_{j=1}^n \sigma_j^{\rm z}$ is a constant of motion, i.e., the Liouvillian is invariant under rotation $U_{\rm Z}={\rm e}^{-\ii\, \alpha Z}$, $U_{\rm Z}\hat{\cal L}(\rho)U_{\rm Z}^\dagger=\hat{\cal L}(U_{\rm Z}\rho U_{\rm Z}^\dagger)$~\cite{foot1}. 
For the analysis of level spacing distribution (LSD) -- a distribution of the differences between the nearest neighbor eigenvalues -- we consider a block of dimension ${n\choose Z}$ of density operator $\rho_m$ with a fixed $Z$ ($m=0$ for NESS and $m \ge 1$ for decay modes) and compute its unfolded~\cite{unfolding} 
spectrum $\{ \lambda_j\}$. 
As a measure of LSD we plot the histogram $p(s)$ of level spacings $s=\lambda_{j+1}-\lambda_j$, and compare it to Poissonian model of uncorrelated levels $p_{\rm Poisson}(s)=\exp(-s)$ or Wigner surmise of GUE $p_{\rm GUE}(s)= \frac{32}{\pi^2}s^2 \exp(-\frac{4}{\pi}s^2)$ \cite{mehta}.

{\em Solvable open spin chains.--} 
We study three instances of qualitatively different non-equilibrium solvable systems: a quadratic non-interacting one, a non-quadratic non-interacting one, and an interacting system, thus enabling us to explore a full range of complexity of solvable non-equilibrium systems.

\begin{figure}[ht!]
\centerline{\includegraphics[angle=-90,width=0.42\textwidth]{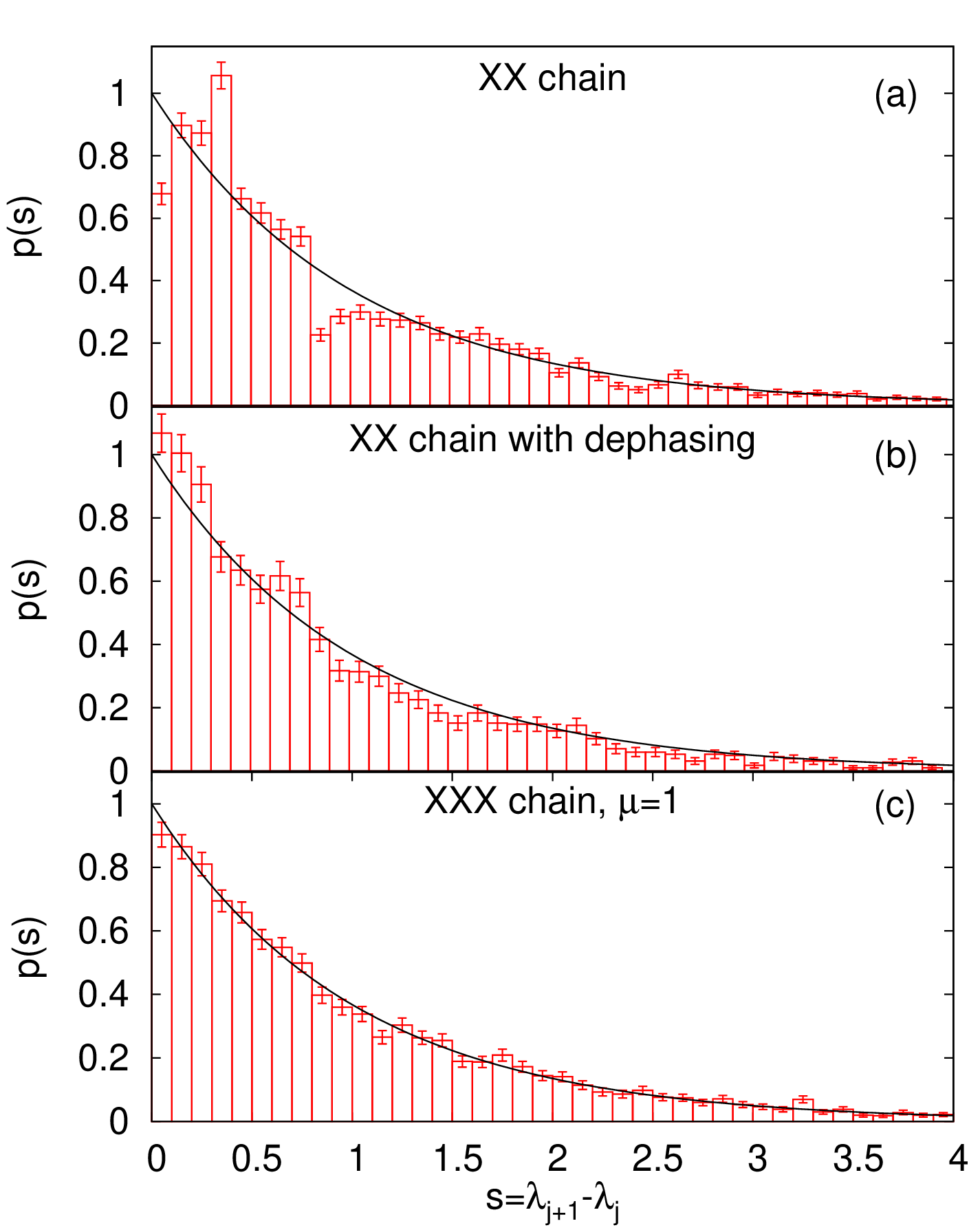}}
\caption{(Color online) Level spacing statistics for NESSs of solvable non-equilibrium systems. a) XX chain ($n=16$, $Z=10$), b) XX chain with dephasing of strength $\gamma=1$ ($n=14$, $Z=7$), c) XXX chain with maximal driving $\mu=1$ ($n=20$, $Z=5$, $\Delta=1$). Cases (a) and (b) are for $\Gamma=1$, $\mu=0.2$, $\bar{\mu}=0.3$, while (c) is for $\Gamma=0.1$, $\mu=1$, $\bar{\mu}=0$.}
\label{fig:lsd-int}
\end{figure}

Perhaps the simplest solvable non-equilibrium model is a boundary driven $XX$ spin chain without dephasing ($\Delta=0$, $b_j=0$, $\gamma=0$). Using Jordan-Wigner transformation the Liouvillian becomes quadratic in fermionic operators and can be readily diagonalized~\cite{3Q}. The system is ballistic due to a non-interaction of fermionic normal modes. We calculate the NESS using a compact matrix product operator form of $\rho_0$ with matrices of fixed dimension $4$~\cite{MPO4}. For $\bar{\mu}=0$ the non-equilibrium XX chain has a parity symmetry $P=XR$, where $X=\prod_{j=1}^n \sigma_j^{\rm x}$ and $R$ is a left-right reflection $R\ket{s_1,s_2,\ldots,s_n}:=\ket{s_n,\ldots,s_2,s_1}$, $s_j\in\{\uparrow,\downarrow\}$, as well as an additional antiunitary symmetry $T=Z_{2} K$, where $K$ is a complex conjugation and $Z_2=\prod_{j=1}^{n/2} \sigma_{2j}^{\rm z}$. To remove these two symmetries we use a nonzero $\bar{\mu}=0.3$. In Fig.~\ref{fig:lsd-int}a the LSD is shown for $n=16$ in the sector with total magnetization $Z=10$. Small deviation from Poissonian statistics can be atributed to finite-size effects.

The next solvable model that we consider is the XX chain ($\Delta=0$, $b_j=0$) with nonzero dephasing for which the system becomes diffusive. The dephasing term $\hat{\cal L}^{\rm deph}$ is quartic in fermionic operators, nevertheless, the NESS can be explicitly written~\cite{dephasing} in powers of the driving $\mu$ due to a closing hierarchy of correlation functions~\cite{eisler}. Nonzero dephasing removes the antiunitary symmetry $T$ while nonzero $\bar{\mu}=0.3$ breaks the parity $P$. We can see in Fig.~\ref{fig:lsd-int}b that the LSD agrees, within statistical fluctuations, with the Poissonian statistics.   

The last and least trivial solvable non-equilibrium case is the $XXZ$ model ($b_j=0$) at maximal driving $\mu=1,\, \bar{\mu}=0$ where $\rho_0$ can be written in terms of an infinite rank matrix product ansatz~\cite{new}. 
As one can see in Fig.~\ref{fig:lsd-int}c the LSD is again Poissonian which suggests existence of Bethe equations for $\{\lambda_j\}$ and Bethe ansatz form of eigenvectors of NESS $\rho_0$ \cite{pip13}.

\begin{figure}[ht!]
\centerline{\includegraphics[angle=-90,width=0.41\textwidth]{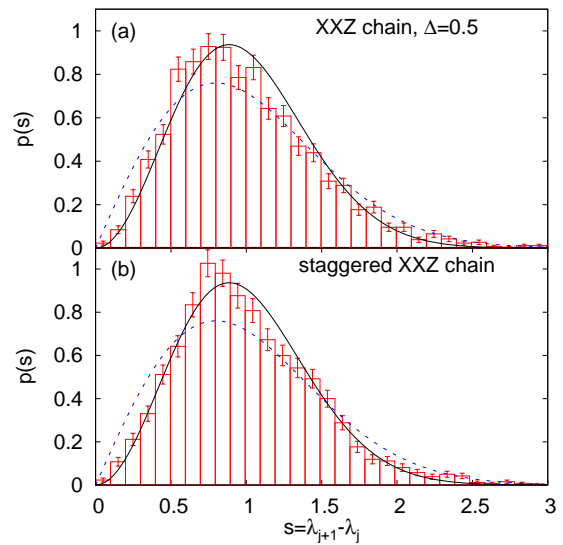}}
\caption{(Color online) Level spacing statistics for NESSs of non-solvable systems. a) XXZ chain with $\Delta=0.5$ ($\Gamma=1$, $\mu=0.2$, $\bar{\mu}=0.3$). b) XXZ chain with $\Delta=0.5$ in a staggered field ($\mu=0.1, \bar{\mu}=0$, $\Gamma=1$). Both cases are for $n=14$ in the sector with $Z=7$. Full black curve is the Wigner surmise for GUE, dotted blue curve is for GOE.} 
\label{fig:lsd-cha}
\end{figure}

\begin{figure}[ht!]
\centerline{\includegraphics[width=0.41\textwidth]{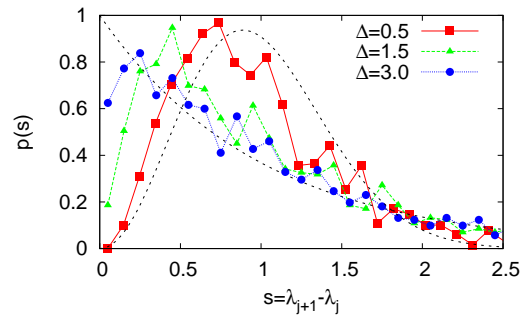}}
\caption{(Color online) Level spacing statistics for NESSs of XXZ chain at different $\Delta$. Increasing $\Delta$ at fixed size $n$, LSD becomes increasingly Poisson-like. All for $n=13$ and the sector with $Z=7$, $\mu=0.2, \bar{\mu}=0.3$, $\Gamma=1$. Dotted black curves are Poissonian and GUE statistics.} 
\label{fig:XXZ}
\end{figure}

{\em Non-solvable open spin chains.--}
Here we consider two instances of a driven open $XXZ$ chain without dephasing. The $XXZ$ model without magnetic field ($b_j=0$) is solvable via Bethe ansatz in its closed-system formulation \cite{kbi93}, however, so-far it has evaded all attempts at finding a non-equilibrium solution (at non-maximal $\mu<1$). Second example is for the $XXZ$ chain in a staggered magnetic field for which the Hamiltonian is quantum chaotic.

For the $XXZ$ model without magnetic field ($b_j=0$, $\gamma=0$) we break parity $P$ by using $\bar{\mu}=0.3$ (antiunitary $T$ is broken by $\Delta$). For nonzero and non-maximal driving $\mu$, non-equilibrium exact solutions are not known. The LSD of NESS, shown in Fig.~\ref{fig:lsd-cha}a for $\Delta=0.5$ and $\mu=0.2$, agrees with the LSD of GUE, describing complex quantum systems without antiunitary symmetry.
 This seems to indicate that such a non-equilibrium system is not solvable. Note that in non-equilibrium states that carry a current, as is the case in all NESSs studied here, we expect the LSD to display GUE statistics and not the one for Gaussian orthogonal ensemble (GOE) irrespective of the symmetry class to which the Hamiltonian belongs (boundary driving will in general break the time-reversal symmetry of $H$). Fixing the system size $n$ and increasing the anisotropy $\Delta$, Fig.~\ref{fig:XXZ}, the LSD becomes increasingly Poisson-like. This might suggest that by increasing $\Delta$ the non-equilibrium $XXZ$ model is perhaps amenable to an exact solution. 
Note that, not surprisingly, the limits $n \to \infty$ and $\Delta \to \infty$ do not commute. Taking a fixed $\Delta$ and increasing $n$ one gets a GUE statistics while fixing $n$ and increasing $\Delta$ one gets a Poissonian statistics. 

\begin{figure}[t!]
\centerline{\includegraphics[width=0.41\textwidth]{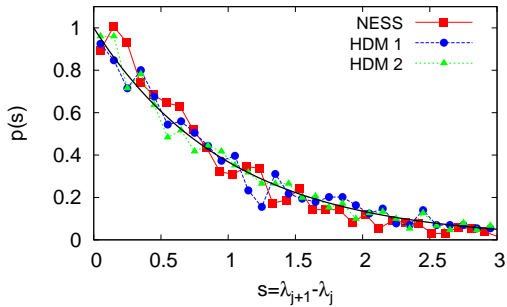}}
\caption{(Color online) Level spacing statistics for two nondegenerate decay modes with the largest real eigenvalue in the XX chain with dephasing $\gamma=1$ ($\mu=0.2, \bar{\mu}=0.3$, $\Gamma=1$, $n=13$, $Z=7$). With red squares we also show the LSD for the NESS (similar data, but different $n$, as in Fig.~\ref{fig:lsd-int}b).} 
\label{fig:XX1decay}
\end{figure}

Switching on an inhomogeneous magnetic field the $XXZ$ model becomes non-integrable even without driving. We shall use a staggered field with period $3$, $b_{3j}=0$, $b_{3j+1}=-1$, $b_{3j+2}=-1/2$, and $\Delta=0.5$, for which the Hamiltonian is quantum chaotic~\cite{thermalization}. Out of equilibrium at $\mu=0.1$ the eigenvalues of the NESS display nice GUE level statistics (Fig.~\ref{fig:lsd-cha}b). 

\begin{figure}[t!]
\centerline{\includegraphics[width=0.41\textwidth]{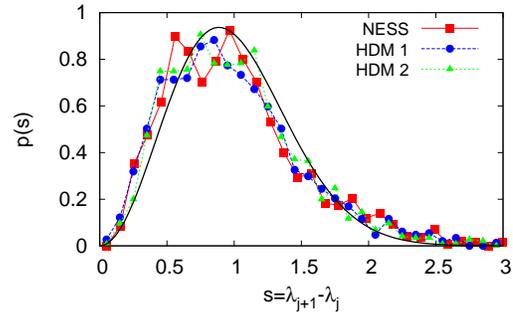}}
\caption{(Color online) Level spacing statistics for two nondegenerate decay modes with the largest real eigenvalue (blue,red) for the XXZ chain in a staggered field. Red points are for the NESS ($\Delta=0.5$, $\mu=0.1, \bar{\mu}=0$, $\Gamma=1$, $n=13$, $Z=7$). Full black curve is the Wigner surmise for GUE.} 
\label{fig:XXZdecay}
\end{figure}

{\em Level spacing distribution of decay modes.--} Apart from quadratic non-equilibrium models solvable by canonical quantization in Liouville space~\cite{3Q} no analytic solution for the decay modes of quantum Liouvillian is known (in the limit $n \to \infty$). 
Calculating two nondegenerate HDMs with the largest real eigenvalue for the XX chain with dephasing, we obtain the LSD shown in Fig.~\ref{fig:XX1decay}. We can see that the statistics is for HDMs (green and blue points) the same as for the NESS (red points). This suggests that the decay modes in the XX chain with dephasing should also be amenable to an analytic calculation. For non-solvable $XXZ$ chain in a staggered field the LSD in Fig.~\ref{fig:XXZdecay} agrees with the distribution for the GUE (HDMs for the $XXZ$ chain without external field, data not shown, also agree with the GUE). Note that small deviations from the GUE theory visible in Fig.~\ref{fig:XXZdecay} are due to a smaller size, $n=13$, than $n=14$ used in Fig.~\ref{fig:lsd-cha}. We can mention that we also calculated LSD for HDMs in a maximally driven $XXZ$ chain ($\mu=1$). Unfortunately, the sizes available ($n \le 13$) do not allow us to make a reliable conclusion about the behavior in the thermodyamic limit and hence to speculate about exact solvability of decay modes. LSD data for HDMs at $n=13$ (not shown) exhibit a Poissonian tail while at the same time showing also some level-repulsion for small spacings.

{\em Conclusion.--} By analyzing Markovian master equations for a variety of boundary driven quantum spin chains we have put forward generalization
of quantum chaos conjecture for non-equilibrium density operators.
We show firm empirical evidence for the correspondence between exact solvability (integrability)  and Poissonian level spacing statistics, on one hand, and between non-integrability and random-matrix statistics, on the other hand.
Consistent results have been found inspecting also other spectral statistics, such as number variance (not shown). Using the generalized conjecture we identify possible new instances of solvable non-equilibrium steady states and decay modes.

Eigenvalues of many-body density operators can be interpreted in terms of occupation probabilities. Statistical fluctuations of these probabilities, discussed here, will influence information-theoretic quantities, like von Neumann entropy, and therefore are of general interest in non-equilibrium quantum physics.

The work has been supported by grant P1-0044 of Slovenian Research Agency (ARRS).


\begin{thebibliography}{10}

\bibitem{mehta} M. L. Mehta, {\em Random Matrices}, 3rd edition, (Elsevier/Academic Press, Amsterdam 2004).

\bibitem{verbaarschot} E.~V.~Shurjak, J.~J.~M.~Verbaarschot, Nucl. Phys. A {\bf 560}, 306 (1993).

\bibitem{guhr} V.~Plerou, P.~Gopikrishnan, B.~Rosenow, L.~A.~Nunes Amaral, T.~Guhr and H.~E.~Stanley, Phys. Rev. E {\bf 65}, 066126 (2002).

\bibitem{berry} M.~V.~Berry,  Ann. Phys. (N.Y.) {\bf 131}, 163 (1981).

\bibitem{CGV81}  G.~Casati,  F.~Valz-Griz and I.~Guarneri, Lett. Nuovo Cimento {\bf 28}, 279 (1980).

\bibitem{BGS84} O.~Bohigas, M.-J.~Giannoni and C.~Schmit, Phys. Rev. Lett. {\bf 52}, 1 (1984).

\bibitem{Muller}   S.~M\" uller, S.~Heusler, P.~Braun, F.~Haake and A.~Altland, Phys. Rev. Lett. {\bf 93}, 014103 (2004) .

\bibitem{BT77} M.~V.~Berry and  M.~Tabor, Proc. R. Soc. A {\bf 356}, 375 (1977).

\bibitem{poilblanc} D.~Poilblanc, T.~Ziman, J.~Bellissard, F.~Mila, and G.~Montambaux, Europhys. Lett. {\bf 22}, 537 (1993).

\bibitem{hjs} H.-J.~St\" ockmann, {\em Quantum Chaos: an introduction}, (Cambridge University Press, Cambridge 1999).

\bibitem{lindblad} V.~Gorini, A.~Kossakowski, and E.~C.~G. Sudarshan, J.~Math.~Phys. {\bf 17}, 821 (1976); G.~Lindblad, Comm. Math. Phys. {\bf 48}, 119 (1976).

\bibitem{note} Since the super-operator $\hat{\cal L}$ is hermiticity preserving, i.e. $\hat{\cal L}(\sigma^\dagger) = (\hat{\cal L}\sigma)^\dagger$ it follows that $\rho_m = (\rho_m)^\dagger$ whenever $\Lambda_m\in\RaR$ and non-degenerate.

\bibitem{marcin} For reduced density operators describing subsystems of closed systems RMT behavior has been found irrespective of integrability of the total Hamiltonian:
M.~Mierzejewski, T.~Prosen, D.~Crivelli and P.~Prelov\v sek, Phys. Rev. Lett. {\bf 110}, 200602 (2013).

\bibitem{blythe} R.~A.~Blythe and M.~R.~Evans,  J. Phys. A Math. Theor. {\bf 40}, R333 (2007).

\bibitem{3Q} T. Prosen, New J. Phys. {\bf 10}, 043026 (2008).

\bibitem{clark} S.~R.~Clark, J.~Prior, M.~J.~Hartmann, D.~Jaksch, and M.~B.~Plenio, New J. Phys. {\bf 12}, 025005 (2010). 

\bibitem{karevski} D.~Karevski and T.~Platini, Phys. Rev. Lett. {\bf 102}, 207207 (2009).

\bibitem{3Qb} T. Prosen and T.~H.~Seligman, J. Phys. A: Math. Theor. {\bf 43}, 392004 (2010).

\bibitem{temme} K. Temme, M.~M.~Wolf, and F.~Verstraete, New. J. Phys. {\bf 14}, 075004 (2012).

\bibitem{horstmann} B.~Hortsmann, J.~I.~Cirac, and G.~Giedke, Phys. Rev. A {\bf 87}, 012108 (2013).

\bibitem{dephasing} M.~\v Znidari\v c, J.~Stat.~Mech. {\bf (2010)}, L05002; M.~\v Znidari\v c, Phys.~Rev.~E {\bf 83}, 011108 (2011).

\bibitem{eisler} V.~Eisler, J.~Stat.~Mech. {\bf (2011)}, P06007.

\bibitem{new} T. Prosen, Phys. Rev. Lett. {\bf 106}, 217206 (2011); Phys. Rev. Lett. {\bf 107} 137201 (2011);  D. Karevski, V. Popkov and G. M. Sch\" utz,  Phys. Rev. Lett. {\bf 110}, 047201 (2013).

\bibitem{pz} T. Prosen and M. \v Znidari\v c, J. Stat. Mech. {\bf  (2009)}, P02035.

\bibitem{robin}  R.~Steinigeweg and J.~Gemmer, Phys. Rev. B. {\bf 80}, 184402 (2009); R.~Steinigeweg, Phys. Rev. E {\bf 84}, 011136 (2011).

\bibitem{markonew} M. \v Znidari\v c,  Phys. Rev. Lett. {\bf 106}, 220601 (2011).

\bibitem{kbi93} V.~E. Korepin, N.~M. Bogoliubov and A.~G. Izergin, ``Quantum inverse scattering method and correlation functions'', (Cambridge Univ. Press, Cambridge 1993).

\bibitem{thermalization} M.~\v Znidari\v c, T.~Prosen, G.~Benenti, G.~Casati, and D.~Rossini, Phys.~Rev.~E {\bf 81}, 051135 (2010).


\bibitem{foot1} This follows due to $U_{\rm Z}\sigma^+ U_{\rm Z}^\dagger={\rm e}^{-\ii 2 \alpha}\sigma^+$ and the fact that the phase of Lindblad operators does not matter.

\bibitem{unfolding}
We chop off $10-20\%$ of eigenvalues at the two ends of the spectrum and perform an {\em unfolding} $\lambda_j \to {\cal N}(\lambda_j)$ where ${\cal N}(\lambda)$ is a low-order polynomial fit to a level counting function, i.e. a number of eigenvalues less than $\lambda$.

\bibitem{MPO4} M.~\v Znidari\v c, J.~Phys.~A {\bf 43}, 415004 (2010).

\bibitem{pip13} T.~Prosen, E.~Ilievski and V.~Popkov,  {\tt  arXiv:1304.7944}, {\em to appear in} New. J. Phys.

\end{thebibliography}
\end{document}